\documentclass{jpp}

\usepackage{graphicx}
\usepackage{natbib}

\ifCUPmtlplainloaded \else
  \checkfont{eurm10}
  \iffontfound
    \IfFileExists{upmath.sty}
      {\typeout{^^JFound AMS Euler Roman fonts on the system,
                   using the 'upmath' package.^^J}%
       \usepackage{upmath}}
      {\typeout{^^JFound AMS Euler Roman fonts on the system, but you
                   dont seem to have the}%
       \typeout{'upmath' package installed. JPP.cls can take advantage
                 of these fonts, if you use 'upmath' package.^^J}%
      }
  \else
  \fi
\fi

\ifCUPmtlplainloaded \else
  \checkfont{msam10}
  \iffontfound
    \IfFileExists{amssymb.sty}
      {\typeout{^^JFound AMS Symbol fonts on the system, using the
                'amssymb' package.^^J}%
       \usepackage{amssymb}%
         \let\leq=\leqslant
         \let\geq=\geqslant
      }{}
  \fi
\fi

\ifCUPmtlplainloaded \else
  \IfFileExists{amsbsy.sty}
    {\typeout{^^JFound the 'amsbsy' package on the system, using it.^^J}%
     \usepackage{amsbsy}}
    {\providecommand\boldsymbol[1]{\mbox{\boldmath $##1$}}}
\fi

\newsavebox{\astrutbox}
\sbox{\astrutbox}{\rule[-5pt]{0pt}{20pt}}

\title[Peculiarities of Surface Plasmons in Quantum Plasmas]{Peculiarities of Surface Plasmons in Quantum Plasmas}

\author[Yu. O. Tyshetskiy, S. V. Vladimirov and R. Kompaneets]%
{Y\ls U\ls R\ls I\ls Y\ns O.\ns T\ls Y\ls S\ls H\ls E\ls T\ls S\ls K\ls I\ls Y$^1$%
  \thanks{Email address for correspondence: yuriy.tyshetskiy@sydney.edu.au},\ns
S.\ns V.\ns V\ls L\ls A\ls D\ls I\ls M\ls I\ls R\ls O\ls V$^{1,2}$\break
\and R.\ns K\ls O\ls M\ls P\ls A\ls N\ls E\ls E\ls T\ls S$^1$}

\affiliation{$^1$School of Physics, University of Sydney,
Sydney, NSW 2006, Australia\\[\affilskip]
$^2$Metamaterials Laboratory, National Research University of Information Technology, Mechanics, and Optics, St Petersburg 199034, Russia}

\pubyear{2013}
\volume{650}
\pagerange{119--126}
\date{?; revised ?; accepted ?. - To be entered by editorial office}
\begin{document}

\maketitle

\begin{abstract}
Surface plasmons (SP) in a semi-bounded quantum plasma with degenerate electrons (e.g., a metal) is considered, and some interesting consequences of electron Pauli blocking for the SP dispersion and temporal attenuation are discussed. In particular, it is demonstrated that a semi-bounded degenerate plasma with a sharp boundary supports \textit{two} types of SP with distinct frequencies and qualitatively different temporal attenuation, in contrast to a non-degenerate plasma that only supports one type of SP~\citep{Guernsey_1969}.
\end{abstract}

\section{Introduction}
The properties of surface plasmons (SP) in bounded plasma structures are defined, among other things, by the dielectric properties of the plasma medium that sustains them. The latter are often (e.g., in metals, for which the electrons are strongly degenerate) significantly affected by the quantum nature of the charge carriers and their interaction in the medium. This can affect the properties of SP in a non-trivial way, via modification of analytic properties of the medium response. In particular, quantum effects (due to Pauli blocking and overlapping wave functions of free charge carriers in the medium~\citep{Eliasson_Shukla_RevModPhys_2011}), when significant, can modify the dispersion, damping~\citep{Tyshetskiy_etal_PoP_2012} and spatial attenuation of SP~\citep{Vladimirov_Kohn_1994} supported by a bounded quantum plasma. 

In this paper, we show another consequence of quantum degeneracy of plasma electrons on SP properties, exemplified by a simple case of SP in a semi-bounded collisionless plasma with degenerate electrons. In particular, we show that such system supports \textit{two} types of SP, with different frequencies and qualitatively different temporal attenuation, in contrast to a case of non-degenerate semi-bounded plasma that only supports one type of SP~\citep{Guernsey_1969}.

\section{Method}
\subsection{Model and assumptions}
We consider a semi-bounded, nonrelativistic collisionless plasma with degenerate mobile electrons ($T_e\ll\epsilon_F$, where $T_e$ is the electron temperature in energy units, $\epsilon_F=\hbar^2(3\pi^2n_e)^{2/3}/2 m_e$ is the electron Fermi energy), and immobile ions; the equilibrium number densities of electrons and ions are equal, $n_{0e}=n_{0i}=n_0$ (quasineutrality). The plasma is assumed to be confined to a region $x<0$, with mirror reflection of plasma particles at the boundary $x=0$ separating the plasma from a vacuum at $x>0$. 

We will look at SPs in the non-retarded limit, when their phase velocity is small compared with the speed of light. In this limit, the SP field is purely electrostatic, hence we can restrain ourselves to considering only electrostatic oscillations in the considered system. Following the discussion of Ref.~\citep{Tyshetskiy_etal_PoP_2012}, we adopt here the quasiclassical kinetic description of plasma electrons in terms of the 1-particle distribution function $f(\mathbf{r,v},t)=f(x,\mathbf{r}_\parallel,v_x,\mathbf{v}_\parallel,t)$~\citep{Vlad_Tysh_UFN_2011} (where $\mathbf{r}_\parallel$ and $\mathbf{v}_\parallel$ are, respectively, the components of $\mathbf{r}$ and $\mathbf{v}$ parallel to the boundary, and $x$ and $v_x$ are the components of $\mathbf{r}$ and $\mathbf{v}$ perpendicular to the boundary), whose evolution is described by the kinetic equation
\begin{equation}
\frac{\partial f}{\partial t} + v_x\frac{\partial f}{\partial x} + \mathbf{v}_\parallel\cdot\frac{\partial f}{\partial\mathbf{r}_\parallel} - \frac{e}{m_e}\left(\frac{\partial\phi}{\partial x}\frac{\partial f}{\partial v_x} + \frac{\partial\phi}{\partial\mathbf{r}_\parallel}\cdot\frac{\partial f}{\partial\mathbf{v}_\parallel}\right) = 0, \label{eq:Vlasov}
\end{equation}
where the electrostatic potential $\phi(x,\mathbf{r}_\parallel,t)$ is defined by the Poisson's equation
\begin{equation}
-\nabla^2\phi = 4\pi e\left[\int{f(\mathbf{r,v},t)d^3\mathbf{v}} - n_0\right]. \label{eq:Poisson}
\end{equation}


In the absence of fields, the equilibrium distribution function of plasma electrons $f_{0}(\mathbf{v})$ is defined by an isotropic Fermi-Dirac distribution, which in the limit $T_e\ll\epsilon_F$ (where $T_e$ is the electron temperature in energy units, and $\epsilon_F=(3\pi^2\hbar^3 n_e)^{2/3}/2m_e$ is the electron Fermi energy) reduces to
\begin{equation}
f_{0}(v)=\frac{2 m_e^3}{(2\pi\hbar)^3}\left\{1 + \exp\left[\frac{m_e v^2/2 - \epsilon_F(n_e)}{T_e}\right] \right\}^{-1} = \frac{2 m_e^3}{(2\pi\hbar)^3} \sigma\left[v_F(n_e)-v\right],  \label{eq:f0}
\end{equation}
where $v_F(n_e)=\sqrt{2\epsilon_F(n_e)/m_e}$ is the electron Fermi velocity, $\sigma(x)$ is the Heaviside step function.

The condition of mirror reflection of plasma electrons off the boundary at $x=0$ implies
\begin{equation}
f(x=0,\mathbf{r}_\parallel,-v_x,\mathbf{v}_\parallel,t) = f(x=0,\mathbf{r}_\parallel,v_x,\mathbf{v}_\parallel,t).  \label{eq:boundary}
\end{equation}

\subsection{Initial value problem}

We now introduce a small initial perturbation $f_p(x,\mathbf{r}_\parallel,v_x,\mathbf{v}_\parallel,t=0)$ to the equilibrium electron distribution function $f_0(v)$, $|f_p(x,\mathbf{r}_\parallel,v_x,\mathbf{v}_\parallel,t=0)|\ll f_0(v)$, and use the kinetic equation (\ref{eq:Vlasov}) to study the resulting evolution of the system's charge density $\rho(x,\mathbf{r}_\parallel,t)=e\left[\int{f(x,\mathbf{r_\parallel,v},t)d^3\mathbf{v}}-n_0\right]$, and hence of the electrostatic potential $\phi(x,\mathbf{r}_\parallel,t)$ defined by (\ref{eq:Poisson}). Introducing the dimensionless variables $\Omega=\omega/\omega_p$, $\mathbf{K}=\mathbf{k}\lambda_F$, $\mathbf{V}=\mathbf{v}/v_F$, $X=x/\lambda_F$, $\mathbf{R}_\parallel=\mathbf{r}_\parallel/\lambda_F$, $\lambda_F=v_F/\sqrt{3}\omega_p$, $\omega_p=(4\pi e^2 n_0/m_e)^{1/2}$, and following Guernsey~\citep{Guernsey_1969}, the solution of the formulated initial value problem for $\rho(X,\mathbf{R}_\parallel,T)$ with the boundary condition (\ref{eq:boundary}) is
\begin{eqnarray}
\rho(X,\mathbf{R}_\parallel,T) &=& en_0\tilde{\rho}(X,\mathbf{R}_\parallel,T), 
\end{eqnarray}
where
\begin{eqnarray}
\tilde{\rho}(X,\mathbf{R}_\parallel,T) &=& \frac{1}{(2\pi)^3}\int_{-\infty}^{+\infty}{dK_x{\ e}^{i K_x X}\int{d^2\mathbf{K}_\parallel}{\ e}^{i\mathbf{K}_\parallel\cdot\mathbf{R}_\parallel}\ \tilde{\rho}_{\mathbf{k}}(T)}, \\
\tilde{\rho}_{\mathbf{K}}(T) &=& \frac{1}{2\pi}\int_{i\sigma-\infty}^{i\sigma+\infty}{\tilde{\rho}(\Omega,\mathbf{K}){\ e}^{-i\Omega T}d\Omega},\ \rm{with\ }\sigma>0. \label{eq:inv_Laplace} 
\end{eqnarray}
The integration in (\ref{eq:inv_Laplace}) is performed in complex $\Omega$ plane along the horizontal contour that lies in the upper half-plane ${\rm Im}(\Omega)=\sigma>0$ above all singularities of the function $\tilde{\rho}(\Omega,\mathbf{K})$. The function $\tilde{\rho}(\Omega,\mathbf{K})$, defined as the Laplace transform of $\tilde\rho_\mathbf{K}(T)$:
\begin{equation}
\tilde{\rho}(\Omega,\mathbf{K}) = \int_0^\infty{\tilde\rho_\mathbf{K}(T)\ e^{i\Omega T} dT}, \label{eq:Laplace}
\end{equation}
is found to be
\begin{eqnarray} 
\tilde{\rho}(\Omega,\mathbf{K}) &=& \frac{i}{\varepsilon(\Omega,K)}\int{d^3\mathbf{V}\frac{G(\mathbf{V,K})}{\Omega-\sqrt{3} \mathbf{K\cdot V}}} \nonumber \\
&+&\frac{iK_\parallel}{2\pi\zeta(\Omega,K_\parallel)}\left[1-\frac{1}{\varepsilon(\Omega,K)}\right]\int_{-\infty}^{+\infty}{\frac{dK_x'}{{K^\prime}^2\ \varepsilon(\Omega,K')}\int{d^3\mathbf{V}\frac{G(\mathbf{V},\mathbf{K^\prime})}{\Omega-\sqrt{3}\mathbf{K'\cdot V}}}},  \label{eq:rho(w,k)} 
\end{eqnarray}
where the Fourier transforms $G(\mathbf{V,K})$ and $G(\mathbf{V,K'})$ of the (dimensionless) initial perturbation are defined by
\begin{eqnarray}
G(\mathbf{V,K}) &=& \int_{-\infty}^{+\infty}{dX {\ e}^{-i K_x X}\  \tilde{g}(X,V_x,\mathbf{V}_\parallel,\mathbf{K}_\parallel)}, 
\end{eqnarray}
with
\begin{eqnarray}
\tilde{g}(X,V_x,\mathbf{V}_\parallel,\mathbf{K}_\parallel) &=& \int{d^2\mathbf{R}_\parallel {\ e}^{-i\mathbf{K_\parallel\cdot R_\parallel}}\  \tilde{f_p}(X,\mathbf{R}_\parallel,V_x,\mathbf{V}_\parallel,0)}, \\
\tilde{f_p}(X,\mathbf{R}_\parallel,V_x,\mathbf{V}_\parallel,0) &=& \frac{v_F^3}{n_0} f_p(X,\mathbf{R}_\parallel,V_x,\mathbf{V}_\parallel,0),
\end{eqnarray}
where $K_\parallel=|\mathbf{K}_\parallel|$, $K=|\mathbf{K}|$, $\mathbf{K}=(K_x,\mathbf{K}_\parallel)$, $K'=|\mathbf{K'}|$, and $\mathbf{K'}=(K_x',\mathbf{K}_\parallel)$.
The functions $\varepsilon(\Omega,K)$ and $\zeta(\Omega,K_\parallel)$ in (\ref{eq:rho(w,k)}) are defined (for ${\rm Im}(\Omega)>0$) as follows:
\begin{eqnarray}
\varepsilon(\Omega,K) &=& 1 - \frac{1}{\sqrt{3}K^2}\int{\frac{\mathbf{K}\cdot\partial\tilde{f}_0(\mathbf{V})/\partial\mathbf{V}}{\Omega-\sqrt{3}\mathbf{K\cdot V}}d^3\mathbf{V}}, \label{eq:epsilon} \\
\zeta(\Omega,K_\parallel) &=& \frac{1}{2} + \frac{K_\parallel}{2\pi}\int_{-\infty}^{+\infty}{\frac{dK_x}{K^2\ \varepsilon(\Omega,K)}}, \label{eq:zeta}
\end{eqnarray} 
with 
\[
\tilde{f}_0(\mathbf{V}) = \frac{v_F^3}{n_0}f_0(\mathbf{V}) = \frac{v_F^3}{n_0}\left.f_0(\mathbf{v})\right|_{\mathbf{v}=v_F\mathbf{V}}. 
\]
For fully degenerate plasma with electron distribution (\ref{eq:f0}), the function $\varepsilon(\Omega,K)$ becomes (for ${\rm Im}(\Omega)>0$)~\citep{Gol'dman_1947,ABR_book}:
\begin{equation}
\varepsilon(\Omega,K) = 1 + \frac{1}{K^2}\left[1-\frac{\Omega}{2\sqrt{3}K}\ln\left(\frac{\Omega+\sqrt{3}K}{\Omega-\sqrt{3}K}\right)\right],\ \ {\rm Im}(\Omega)>0,  \label{eq:epsilon_degenerate}
\end{equation}
where $\ln(z)$ is the principal branch of the complex natural logarithm function.

Note that the solution~(\ref{eq:rho(w,k)}) differs from the corresponding solution of the transformed Vlasov-Poisson system for infinite (unbounded) uniform plasma only in the second term involving $\zeta(\Omega,K_\parallel)$; indeed, this term appears due to the boundary at $x=0$.


The definition (\ref{eq:Laplace}) of the function $\tilde\rho(\Omega,\mathbf{K})$ of complex $\Omega$ has a sense (i.e., the integral in (\ref{eq:Laplace}) converges) only for ${\rm Im}(\Omega)>0$. Yet the long-time evolution of $\tilde{\rho}_\mathbf{k}(T)$ is obtained from (\ref{eq:inv_Laplace}) by displacing the contour of integration in complex $\Omega$ plane from the upper half-plane ${\rm Im}(\Omega)>0$ into the lower half-plane ${\rm Im}(\Omega)\leq0$~\citep{Landau_1946}. This requires the definition of $\tilde\rho(\Omega,\mathbf{K})$ to be extended to the lower half-plane, ${\rm Im}(\Omega)\leq 0$, by analytic continuation of (\ref{eq:rho(w,k)}) from ${\rm Im}(\Omega)>0$ to ${\rm Im}(\Omega)\leq 0$. Hence, the functions
\begin{equation}
I(\Omega,\mathbf{K})\equiv\int{d^3\mathbf{V}\frac{G(\mathbf{V,K})}{\Omega-\sqrt{3}\mathbf{K\cdot V}}},  \label{eq:I}
\end{equation}
$\varepsilon(\Omega,K)$, and $\zeta(\Omega,K_\parallel)$ that make up the function $\tilde\rho(\Omega,\mathbf{K})$, must also be analytically continued into the lower half-plane of complex $\Omega$, thus extending their definition to the whole complex $\Omega$ plane. With thus continued functions, the contributions to the inverse Laplace transform (\ref{eq:inv_Laplace}) are of three sources~\citep{Guernsey_1969}: 
\begin{enumerate}
\item Contributions from the singularities of $I(\Omega,\mathbf{K})$ in the lower half of complex $\Omega$ plane (defined solely by the initial perturbation $G(\mathbf{V,K})$); with some simplifying assumptions about the initial perturbation~\citep{Guernsey_1969} these contributions are damped in a few plasma periods and can be ignored. 
\item Contribution of singularities of $1/\varepsilon(\Omega,K)$ in the lower half of complex $\Omega$ plane, of two types: (i) residues at the poles of $1/\varepsilon(\Omega,K)$, which give the volume plasma oscillations~\citep{Guernsey_1969}, and (ii) integrals along branch cuts (if any) of $1/\varepsilon(\Omega,K)$ in the lower half-plane of complex $\Omega$, which can lead to non-exponential attenuation of the volume plasma oscillations~\citep{Hudson_1962,Krivitskii_Vladimirov_1991}.
\item Contribution into (\ref{eq:inv_Laplace}) of singularities of $1/\zeta(\Omega,K_\parallel)$ in the lower half of complex $\Omega$ plane, of two types: (i) residues at the poles of $1/\zeta(\Omega,K_\parallel)$, corresponding to the surface wave solutions of the initial value problem in the considered system~\citep{Guernsey_1969,Tyshetskiy_etal_PoP_2012}, and (ii) integrals along branch cuts (if any) of $1/\zeta(\Omega,K_\parallel)$ in the lower half-plane of complex $\Omega$. 
\end{enumerate}

\begin{figure}
\centerline{\includegraphics[width=2.7in]{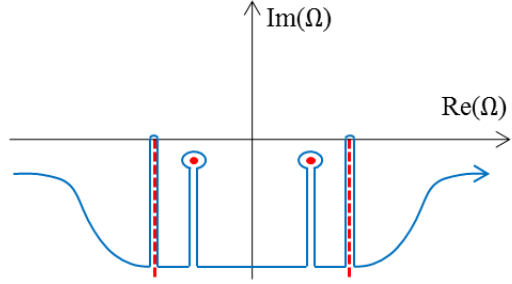}}
\caption{A sketch of the deformed integration path in (\ref{eq:inv_Laplace}) in complex $\Omega$ plane, with contributions of poles (solid circles) and branch cuts (dashed lines) of $1/\zeta(\Omega,K_\parallel)$. The singularities due to $1/\varepsilon(\Omega,K)$ are not shown, but they also contribute to (\ref{eq:inv_Laplace}), yielding volume plasmons.}
\label{fig:sketch_integration_omega}
\end{figure}

Below we consider the latter contributions from poles and branch cuts of $1/\zeta(\Omega,K_\parallel)$ in the lower half-plane of complex $\Omega$, as illustrated in Fig.~\ref{fig:sketch_integration_omega}, and show that they yield two types of electrostatic surface oscillations with different frequencies and qualitatively different temporal attenuation.

\section{Two types of surface oscillations} 
\subsection{Contribution of poles of $1/\zeta(\Omega,K_\parallel)$}
The contribution of poles of $1/\zeta(\Omega,K_\parallel)$ into (\ref{eq:inv_Laplace}) leads to exponentially damped surface oscillations~\citep{Tyshetskiy_etal_PoP_2012}
\begin{equation}
\tilde\rho_{\mathbf{K}}^{\rm (poles)}(T)\propto e^{-|\Gamma_s|T}\cos\left(\Omega_s T\right), \label{eq:rho(T)_poles}
\end{equation}
with frequency $\Omega_s=\omega_s/\omega_p$ and damping rate $\Gamma_s=\gamma_s/\omega_p$ obtained from the dispersion equation $\zeta(\Omega,K_\parallel)=0$.
The frequency asymptotes are
\begin{eqnarray}
\Omega_s &\approx& \frac{1}{\sqrt{2}}\left(1+0.95 K_\parallel\right),\ \rm{for\ }K_\parallel\ll 1  \label{eq:Omega_Kz<<1} \\
\Omega_s &\approx& \sqrt{3} K_\parallel \left(1+2\exp\left[-2-4K_\parallel^2\right]\right),\ \rm{for\ }K_\parallel\gg 1\ (\rm zero\ sound) \label{eq:dispersion_Kz>>1}
\end{eqnarray}
The absolute value of the damping rate is a nonmonotonic function of $K_\parallel$. At small $K_\parallel$, it increases linearly with $K_\parallel$,
\begin{equation}
\left|\Gamma_s(K_\parallel)\right|\approx 2.1\sqrt{3}\cdot 10^{-2} K_\parallel,  \label{eq:Gamma_Kz<<1}
\end{equation}
reaches maximum $|\Gamma_s|\approx 6.2\cdot 10^{-3}$ at $K_\parallel\approx 0.4$, and then quickly decreases at $K_\parallel>0.4$. Since the maximum growth rate is small, the surface oscillations due to the poles of $1/\zeta(\Omega,K_\parallel)$ are weakly damped at all wavelengths~\citep{Tyshetskiy_etal_PoP_2012}.


\subsection{Contribution of branch cuts of $1/\zeta(\Omega,K_\parallel)$ \label{sec:cuts}}
For degenerate plasma, the analytically continued function $\zeta(\Omega, K_\parallel)$ has two branching points on the real axis of the complex $\Omega$ plane at $\Omega=\pm\Omega_v(K_\parallel)$, where $\Omega_v(K_\parallel)\in\mathbb{R}$ is the solution of equation
\begin{equation}
\varepsilon(\Omega,K_\parallel)=\left.\varepsilon(\Omega,K)\right|_{K_x=0}=0,  \label{eq:Omega_v}
\end{equation}
with the corresponding branch cuts going down into the ${\rm Im}(\Omega)<0$ part of the complex $\Omega$ plane, as schematically shown in Fig.~\ref{fig:sketch_integration_omega} (see Appendix~\ref{app:branching}). Let us consider the contribution of the integration along these branch cuts into the inverse Laplace transform (\ref{eq:inv_Laplace}). The branching points lie above the poles $\Omega_s-i|\Gamma_s|$ of $1/\zeta$ (since the latter lie below the real axis of the $\Omega$ plane), therefore we can expect the contribution of the integration along the branch cuts into (\ref{eq:inv_Laplace}) to be at least as important as the contribution of the poles, if not to exceed it.

\begin{figure}
\centerline{\includegraphics[width=3.5in]{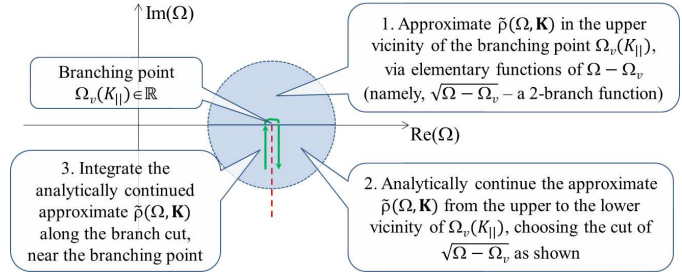}}
\caption{Evaluation scheme for the integrals along branch cuts, as outlined in Sec.~\ref{sec:cuts}.}
\label{fig:anal_cont}
\end{figure}

At large times $T\gg 1$, the main contribution into the integrals along the branch cuts comes from the small vicinity of the branching points, so it suffices to approximate the second term of (\ref{eq:rho(w,k)}) near the branching points in the lower semiplane of complex $\Omega$. This can be done in two steps:
\begin{enumerate}
\item approximate $\tilde\rho(\Omega,\mathbf{K})$ defined by (\ref{eq:rho(w,k)}) in the upper vicinities of the branching points, in terms of elementary functions; the approximate function should have the same branching points as the original one;
\item analytically continue these approximations into the lower vicinities of the branching points, choosing the branch cuts to go down.
\end{enumerate}
Then we can perform the integration of thus obtained approximations along the branch cuts in the vicinity of the branching points. This scheme is sketched in Fig.~\ref{fig:anal_cont}

The function (\ref{eq:rho(w,k)}) in the upper vicinity of the right branching point $\Omega=+\Omega_v(K_\parallel)$ can be approximated as~(see Appendix~\ref{app:rho_approximate})
\begin{eqnarray}
\tilde\rho(\Omega,\mathbf{K})&\approx& i\frac{I(\Omega_v,\mathbf{K})}{\left.\varepsilon(\Omega,K)\right|_{\Omega\approx\Omega_v}} \nonumber \\
&+&i\left[1-\frac{1}{\left.\varepsilon(\Omega,K)\right|_{\Omega\approx\Omega_v}}\right]\frac{\left.I(\Omega_v,\mathbf{K})\right|_{K_x=0}}{1+\sqrt{\alpha(K_\parallel)\beta(K_\parallel)}K_\parallel\sqrt{\Omega-\Omega_v}}, \label{eq:rho_approx}
\end{eqnarray}
where $I(\Omega,\mathbf{K})$ is defined in Eq.~(\ref{eq:I}) and is assumed to vary slowly near the point $\Omega=\Omega_v$ and to not have branch cuts. The expansion of $\varepsilon(\Omega,K)$ near $\Omega=\Omega_v$ is
\begin{equation}
\left.\varepsilon(\Omega,K)\right|_{\Omega\approx\Omega_v} = \varepsilon(\Omega_v,K) + \beta(K)(\Omega-\Omega_v) + O\left[(\Omega-\Omega_v)^2\right], \label{eq:epsilon_approx}
\end{equation}
with $\alpha(K_\parallel)$, $\beta(K_\parallel)$ and $\beta(K)$ defined in Eqs~(\ref{eq:alpha})--(\ref{eq:beta(K)}) of Appendix~\ref{app:rho_approximate} (note that in general $\beta(K)\neq\beta(K_\parallel)$, since $K\neq K_\parallel$). The approximation (\ref{eq:rho_approx}) with (\ref{eq:epsilon_approx}) is expressed in terms of elementary functions of $\Omega-\Omega_v$, which can be analytically continued into the lower vicinity ${\rm Im}(\Omega)<0$ of the branching point $\Omega=\Omega_v$. When doing so, the complex function $\sqrt{\Omega-\Omega_v}$ should be defined so that its branch cut goes vertically down from its branching point $\Omega=\Omega_v$. This branch cut can be parametrized as
\[
\Omega_{\rm cut}=+\Omega_v(K_\parallel) - i\eta,\ \ \eta\geq 0.
\]

The integral around the branch cut in the vicinity of the right branching point $\Omega=+\Omega_v$ is (the $(+)$ superscript denotes the right branch cut)
\begin{eqnarray}
\tilde\rho^{(+)}_\mathbf{K}(T) = \frac{i}{2\pi}e^{-i\Omega_vT}\int_0^\infty{d\eta e^{-\eta T}\left[\tilde\rho^{(+)}_L(\Omega_v-i\eta,\mathbf{K}) - \tilde\rho^{(+)}_R(\Omega_v-i\eta,\mathbf{K})\right]}, \label{eq:rho+_gen}
\end{eqnarray}
where $\tilde\rho^{(+)}_{L,R}(\Omega,\mathbf{K})$ are the left and right branches of the analytic continuation of (\ref{eq:rho_approx}) into the lower semiplane ${\rm Im}(\Omega)<0$.

We first assume that the function $\varepsilon(\Omega,K)$, analytically continued to ${\rm Im}(\Omega)\leq0$, does not have branching points in some (perhaps small) vicinity of the point $\Omega=\Omega_v$; hence we have 
\[
\varepsilon_L(\Omega,K) = \varepsilon_R(\Omega,K)
\]
near $\Omega=\Omega_v$. Then the only function in $\tilde\rho^{(+)}_{L,R}(\Omega,\mathbf{K})$ with a cut is $\sqrt{\Omega-\Omega_v}$. Choosing its cut as specified above, we have
\begin{eqnarray}
\left(\sqrt{\Omega-\Omega_v}\right)_R = -\left(\sqrt{\Omega-\Omega_v}\right)_L = \sqrt{\left|\Omega-\Omega_v\right|}e^{-i\pi/4} = \sqrt{|\eta|}e^{-i\pi/4}.  \label{eq:sqrt}
\end{eqnarray}

Using (\ref{eq:rho_approx}), (\ref{eq:sqrt}) and (\ref{eq:epsilon_approx}), from (\ref{eq:rho+_gen}) we obtain
\begin{eqnarray}
\tilde\rho^{(+)}_\mathbf{K}(T) &=& \frac{ie^{i\pi/4}}{\pi}\sqrt{\alpha(K_\parallel)\beta(K_\parallel)}K_\parallel \left.I(\Omega_v,\mathbf{K})\right|_{K_x=0} e^{-i\Omega_v T} \nonumber \\
&\times& \int_0^\infty{d\eta \frac{e^{-\eta T} \sqrt{\eta}}{1 + i\eta\ \alpha(K_\parallel)\beta(K_\parallel)K_\parallel^2}\left[1-\frac{1}{\varepsilon(\Omega_v,K) - i\eta\beta(K) + O(\eta^2)}\right]}.  \label{eq:rho(T)}
\end{eqnarray}
The long-time asymptote of $\tilde\rho^{(+)}_\mathbf{K}(T)$ depends on whether $\varepsilon(\Omega_v,K)$ tends to zero or not, which in turn depends on the value of $K_x$. Below we consider the two cases: (i) $K_x\neq0$, so that $\varepsilon(\Omega_v,K)\neq0$, and (ii) $K_x\to0$, so that $\varepsilon(\Omega_v,K)\to 0$ [since $\Omega_v$ is the root of (\ref{eq:Omega_v})]. In these cases, (\ref{eq:rho(T)}) gives, respectively:
\begin{eqnarray}
\tilde\rho^{(+)}_{\mathbf{K}}(T)&\propto&\frac{e^{-i\Omega_vT}}{T^{3/2}} + O(T^{-5/2}),\ \ K_x\neq0, \label{eq:rho+_Kx!=0} \\
\tilde\rho^{(+)}_{\mathbf{K}}(T)&\propto&\frac{e^{-i\Omega_vT}}{T^{1/2}} + O(T^{-3/2}),\ \ K_x\to0.  \label{eq:rho+_Kx=0}
\end{eqnarray}
Similarly for the contribution of the left branch cut, $\Omega=-\Omega_v-i\eta,\ \eta\geq0$, we have $\tilde\rho^{(-)}_{\mathbf{K}}(T)$ given by Eqs~(\ref{eq:rho+_Kx!=0})--(\ref{eq:rho+_Kx=0}) with $\Omega_v$ replaced with $-\Omega_v$. The total contribution of both branch cuts $\tilde\rho^{\rm (cuts)}_{\mathbf{K}}(T)=\tilde\rho^{(+)}_{\mathbf{K}}(T)+\tilde\rho^{(-)}_{\mathbf{K}}(T)$ is then
\begin{eqnarray}
\tilde\rho^{\rm (cuts)}_{\mathbf{K}}(T)&\propto&\frac{\cos\left(\Omega_vT\right)}{T^{3/2}} + O(T^{-5/2}),\ \ K_x\neq0, \label{eq:rho(T)_cuts_Kx!=0} \\
\tilde\rho^{\rm (cuts)}_{\mathbf{K}}(T)&\propto&\frac{\cos\left(\Omega_vT\right)}{T^{1/2}} + O(T^{-3/2}),\ \ K_x\to0.  \label{eq:rho(T)_cuts_Kx=0}
\end{eqnarray}
Note that the frequency of these oscillations is equal to the frequency of volume plasma waves with the same wavelength, $K=K_\parallel$, and thus exceeds the frequency $\Omega_s$ of the surface oscillations (\ref{eq:rho(T)_poles}) due to the poles of $1/\zeta$.

Above we have assumed that the function $\varepsilon(\Omega,K)$, analytically continued to ${\rm Im}(\Omega)\leq0$, does not branch in at least some vicinity of the branching point $\Omega_v$ of $\zeta(\Omega,K_\parallel)$. However, this assumption is violated in a special case considered below. Indeed, the function $\varepsilon(\Omega,K)$ itself, when analytically continued into ${\rm Im}(\Omega)\leq0$, has branching points at $\Omega=\pm\sqrt{3}K\in\mathbb{R}$, with the branch cuts going down~\citep{Vlad_Tysh_UFN_2011}. Thus for $K_x\to\sqrt{\Omega_v^2(K_\parallel)/3 - K_\parallel^2}$ the branching points of $\varepsilon(\Omega,K)$ merge with the branching points of $\zeta(\Omega,K_\parallel)$, and their respective branch cuts merge at least in some lower vicinity of the coinciding branching points. In this case $\varepsilon_L(\Omega,K)\neq\varepsilon_R(\Omega,K)$, and the above calculation is modified; instead, we have for ${\rm Im}(\Omega)\leq0$
\begin{eqnarray}
\varepsilon_R(\Omega,K) &=& 1+\frac{1}{K^2}\left[1-\frac{\Omega}{2\sqrt{3}K}\log\left(\frac{\Omega+\sqrt{3}K}{\Omega-\sqrt{3}K}\right)\right], \\
\varepsilon_L(\Omega,K) &=& \varepsilon_R(\Omega,K) + \frac{i\pi\Omega}{\sqrt{3}K^3}
\end{eqnarray}
(we still assume that $I(\Omega,\mathbf{K})$ does not have branching points). Then, after some calculation, we obtain for $\tilde\rho^{(+)}_L(\Omega_v-i\eta,\mathbf{K}) - \tilde\rho^{(+)}_R(\Omega_v-i\eta,\mathbf{K})$ in (\ref{eq:rho+_gen}) in this case:
\begin{equation}
\tilde\rho^{(+)}_L(\Omega_v-i\eta,\mathbf{K}) - \tilde\rho^{(+)}_R(\Omega_v-i\eta,\mathbf{K}) = -i\left.\frac{I(\Omega_v,\mathbf{K})}{\varepsilon_R(\Omega_v,K)}\right|_{K_x=\sqrt{\Omega_v^2/3-K_\parallel^2}} + O\left(|\eta|^{1/2}\right).
\end{equation}
Carrying out the integration in (\ref{eq:rho+_gen}) and adding the similar contribution of the left cut, we finally obtain for the contribution of branch cuts in this special case:
\begin{equation}
\tilde\rho^{\rm (cuts)}_{\mathbf{K}}(T)\propto\frac{\cos(\Omega_v T)}{T} + O\left(T^{-3/2}\right),\ \ {\rm for\ } K_x\to\sqrt{\Omega_v^2/3-K_\parallel^2},  \label{eq:rho(T)_cuts_Kx=Kxc}
\end{equation}
which has the same frequency as (\ref{eq:rho(T)_cuts_Kx!=0})--(\ref{eq:rho(T)_cuts_Kx=0}), but a different temporal attenuation exponent.

\section{Discussion}
We thus see that our system supports two distinct types of surface oscillations, with different frequencies and temporal attenuation: 
\begin{enumerate}
\item exponentially damped surface oscillations (\ref{eq:rho(T)_poles}) with frequency $\Omega_s(K_\parallel)$, due to the poles of $\tilde\rho(\Omega,\mathbf{K})$~\citep{Tyshetskiy_etal_PoP_2012};
\item power-law attenuated surface oscillations (\ref{eq:rho(T)_cuts_Kx!=0}), (\ref{eq:rho(T)_cuts_Kx=0}), and (\ref{eq:rho(T)_cuts_Kx=Kxc}) with frequency $\Omega_v(K_\parallel)>\Omega_s(K_\parallel)$, due to the branch cuts of $\tilde\rho(\Omega,\mathbf{K})$. Since the power-law attenuation is slower than the exponential attenuation, these oscillations should become dominant at large times, and should theoretically become observable. However, by the time this happens, the level of the oscillations may well become too low due to the attenuation, making a practical detection of this branch of oscillations rather difficult. 

It is interesting to note that, as seen from (\ref{eq:rho(T)_cuts_Kx!=0}), (\ref{eq:rho(T)_cuts_Kx=0}), and (\ref{eq:rho(T)_cuts_Kx=Kxc}), different $K_x$ components in the wave packet making up the surface oscillation of this type is attenuated at different rate. Since $T^{-1/2}$ decays slower than $T^{-1}$ or $T^{-3/2}$, the small-$K_x$ part of the wave packet becomes dominant over the large-$K_x$ part at large times, which corresponds to penetration of the perturbation away from the surface and deeper into plasma.
\end{enumerate}  

The presented analysis relies on several assumptions discussed in detail in Ref.~\citep{Tyshetskiy_etal_PoP_2012}, of which perhaps the most critical ones are the assumptions of collisionless plasma and of the sharp perfectly reflecting boundary confining the plasma. While relaxing the former assumption does not change the results qualitatively~\citep{Tyshetskiy_etal_PoP_2012(2)}, relaxing the latter assumption may change the results dramatically. Firstly, the smooth boundary leads to a new resonant damping of surface oscillations, dramatically increasing the exponential damping rate $|\Gamma_s|$ in (\ref{eq:rho(T)_poles})~\citep{Marklund_etal_new_quantum_limits}. Secondly, allowing for boundary smoothness (with a simultaneous account for the quantum tunneling, as they both have the same spatial scales) should change the analytic properties of $\tilde\rho(\Omega,\mathbf{K})$ in the lower semiplane ${\rm Im}(\Omega)<0$ of complex $\Omega$ plane, and thus may change its branch cuts and their contribution into (\ref{eq:inv_Laplace}). This, however, is a more elaborate problem, which is beyond the scope of this paper and is left for future work.

\appendix
\section{Branching points of $\zeta(\Omega,K_\parallel)$ \label{app:branching}}
Let us show that the function $\zeta(\Omega,K_\parallel)$, analytically continued into ${\rm Im}(\Omega)\leq0$, has branching points at $\Omega=\pm\Omega_v(K_\parallel)\in\mathbb{R}$, with the corresponding branch cuts going down from these two points. We start from $\zeta(\Omega,K_\parallel)$ defined by Eq.~(\ref{eq:zeta}) for ${\rm Im}(\Omega)>0$, and then continuously change ${\rm Im}(\Omega)$ to negative values. In this process, we must consider how the singularities of the function $\left[K^2\varepsilon(\Omega,K)\right]^{-1}$ under the integral in (\ref{eq:zeta}) change in the complex $K_x$ plane~\citep{Tyshetskiy_etal_PoP_2012}. These singularities are:
\begin{enumerate}
\item Branch cuts of the complex square root $\sqrt{K_x^2+K_\parallel^2}$, defined by two parametric equations:
\begin{equation}
K_x = \pm i \sqrt{K_\parallel^2 + \tau},\ {\rm with\ }K_\parallel>0,\ \tau\in[0,+\infty). \label{eq:sqrt_cuts}
\end{equation}
\item Branch cut of the complex logarithm in (\ref{eq:epsilon_degenerate}), taken along the negative real axis of the argument $(\Omega+\sqrt{3}K)/(\Omega-\sqrt{3}K)$. This branch cut maps into two branch cuts of $\left[K^2\ \varepsilon(\Omega,K)\right]^{-1}$ in the complex $K_x$ plane, given by two parametric equations:
\begin{equation}
K_x = \pm i \sqrt{\frac{\Omega^2}{3}\left(\frac{\tau+1}{\tau-1}\right)^2 - K_\parallel^2},\ {\rm with\ }\Omega\in\mathbb{C},\ K_\parallel>0,\ \tau\in[0,+\infty).  \label{eq:log_cuts}
\end{equation}
\item Two poles $K_x=\pm i K_\parallel$ ($K_\parallel>0$) at the roots of $K^2=0$, lying symmetrically above and below the real axis of the complex $K_x$ plane.
\item Two poles $\pm K_x^{r}\in\mathbb{C}$ at the roots of $\varepsilon(\Omega,K)=0$. Note that for any $K\in\mathbb{R}$, $\varepsilon(\Omega,K)=0$ does not have roots with ${\rm Im}(\Omega)>0$, if the plasma equilibrium is stable~\citep{Penrose_1960}, which is the case considered here; therefore, for any ${\rm Im}(\Omega)>0$ the poles $\pm K_x^{r}$ are located \textit{away} from the real axis of the complex $K_x$ plane, and thus do not lie on the integration contour in (\ref{eq:zeta}). 
\end{enumerate}
\begin{figure}
\centerline{\includegraphics[width=2.2in]{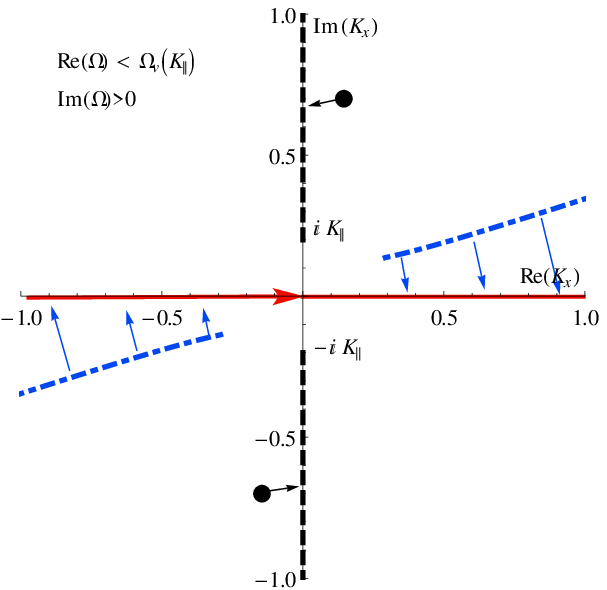}\ \ \ \includegraphics[width=2.2in]{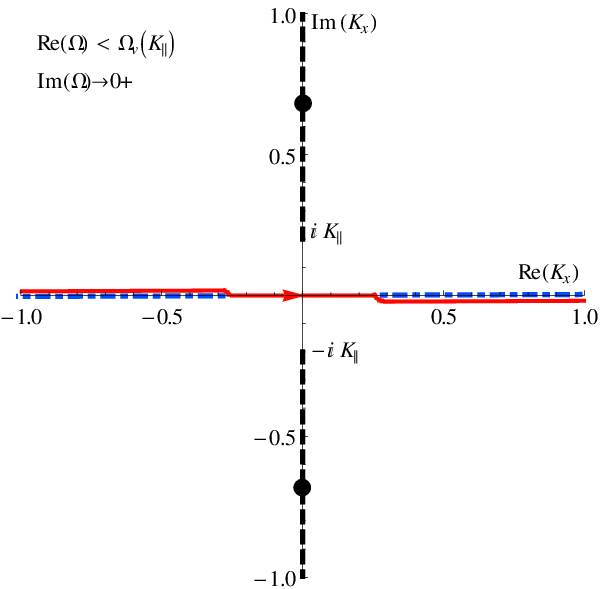}} 
\centerline{\includegraphics[width=2.2in]{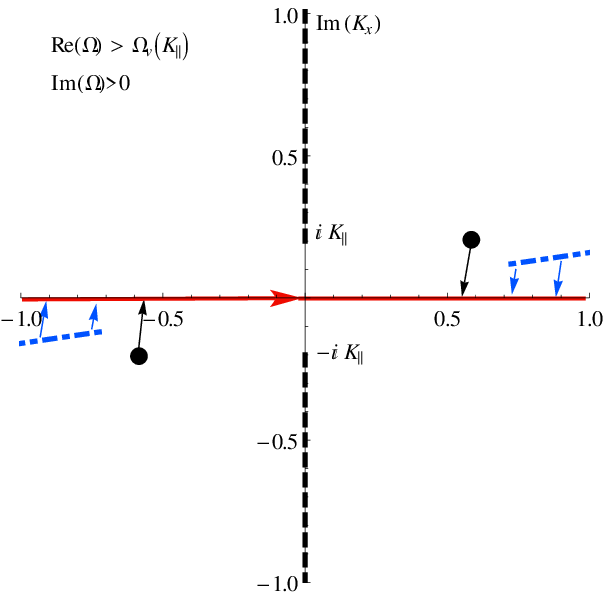}\ \ \ \includegraphics[width=2.2in]{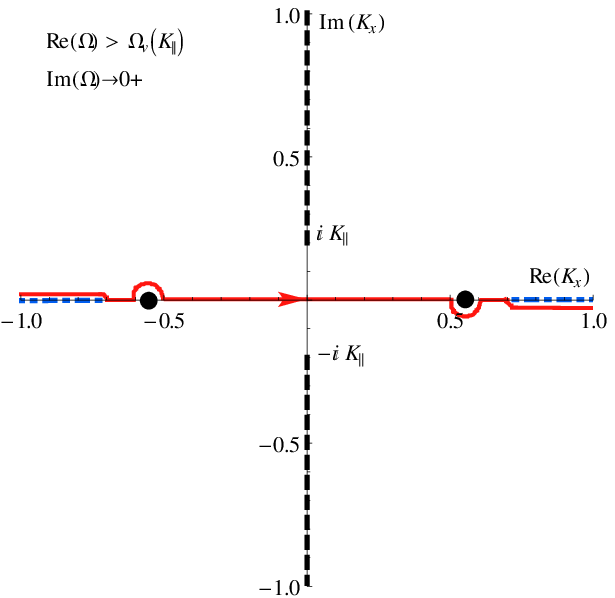}}
\caption{\label{fig:Kx_plane} (Color online) Singularities of the function $\left[K^2\ \varepsilon(\Omega,K)\right]^{-1}$ in the complex $K_x$ plane, and their modification from ${\rm Im}(\Omega)>0$ (left panels) to the limit ${\rm Im}(\Omega)\rightarrow0+$ (right panels), for $0<{\rm Re}(\Omega)<\Omega_v(K_\parallel)$ (upper row) and ${\rm Re}(\Omega)>\Omega_v(K_\parallel)>0$ (lower row), with $\Omega_v(K_\parallel)$ defined from (\ref{eq:Omega_v}). The branch cuts (\ref{eq:sqrt_cuts}) and (\ref{eq:log_cuts}) are shown with the black dashed lines and the blue dot-dashed lines, respectively. The poles $\pm K_x^{r}$ where $\varepsilon(\Omega,K)=0$ are shown with the filled circles. The arrows show the direction of motion of the singularities when ${\rm Im}(\Omega)\rightarrow0+$. The contour of $K_x$ integration in (\ref{eq:zeta}) is shown with the solid red line.}
\end{figure}

In the process of analytic continuation, as ${\rm Im}(\Omega)\to 0+$, these singularities deform/move in the complex $K_x$ plane, as shown in Fig.~\ref{fig:Kx_plane}. We have the following cases: (i) $|{\rm Re}(\Omega)|<|\Omega_v(K_\parallel)|$, and (ii) $|{\rm Re}(\Omega)|>|\Omega_v(K_\parallel)|$, where $\pm\Omega_v(K_\parallel)\in\mathbb{R}$ are defined by the equation~(\ref{eq:Omega_v}). The difference between these two cases is that for $|{\rm Re}(\Omega)|>|\Omega_v(K_\parallel)|$, the poles $\pm K_x^{r}$ cross the real axis in $K_x$ plane and deform the integration contour when ${\rm Im}(\Omega)\to 0+$ and beyond to negative values, while for $|{\rm Re}(\Omega)|<|\Omega_v(K_\parallel)|$ they do not. Thus we have that in these two cases, the integration contours in the function $\zeta(\Omega,K_\parallel)$ continued to ${\rm Im}(\Omega)<0$ are \textit{different}, and thus the values of $\zeta(\Omega,K_\parallel)$ in the lower semiplane of complex $\Omega$ are also different for $|{\rm Re}(\Omega)|<|\Omega_v(K_\parallel)|$ and for $|{\rm Re}(\Omega)|>|\Omega_v(K_\parallel)|$. Hence, the points $\Omega=\pm\Omega_v(K_\parallel)\in\mathbb{R}$ separating these two cases must necessarily be the branching points of $\zeta(\Omega,K_\parallel)$ in ${\rm Im}(\Omega)\leq0$, with branch cuts (separating the different values of the analytically continued $\zeta$) going down into the lower semiplane of complex $\Omega$. At the branching points $\pm\Omega_v(K_\parallel)$, the function $\zeta(\Omega,K_\parallel)$ has a singularity~\citep{Tyshetskiy_etal_PoP_2012}.

\section{Approximation of $\tilde\rho(\Omega,\mathbf{K})$ in the upper vicinity of the branching point $\Omega=+\Omega_v(K_\parallel)$ \label{app:rho_approximate}}
The function $\tilde\rho(\Omega,\mathbf{K})$ defined in (\ref{eq:rho(w,k)}) contains integrals of the form
\begin{equation}
\int_{-\infty}^{+\infty}\frac{dK_x}{K^2 \varepsilon(\Omega,K)}A(\Omega,K) = \frac{1}{K_\parallel^2}\int_{-\infty}^{+\infty}\frac{A(\Omega,K) dK_x}{\left(1+K_x^2/K_\parallel^2\right) \varepsilon(\Omega,K)},  \label{eq:int_Kx}
\end{equation}
which need to be approximated in the upper vicinity of the branching point $\Omega=\Omega_v$. The main contribution into the integrals (\ref{eq:int_Kx}) is from the vicinity of $K_x=0$. The expansion of $\varepsilon(\Omega,K)$ near $\Omega=\Omega_v$, $K_x=0$ under the integral is
\begin{eqnarray}
\varepsilon(\Omega,K)=\left(\Omega-\Omega_v\right)\beta(K_\parallel) + \alpha(K_\parallel)K_x^2 + O\left[\left(\Omega-\Omega_v\right)^2\right] + O\left(K_x^4\right),
\end{eqnarray}
where
\begin{eqnarray}
\alpha(K_\parallel) &=& \frac{1}{2}\left.\varepsilon^{\prime\prime}_{Kx,Kx}(\Omega,K)\right|_{K_x=0,\ \Omega=\Omega_v}, \label{eq:alpha}\\
\beta(K_\parallel) &=& \left.\varepsilon^\prime_\Omega(\Omega,K)\right|_{K_x=0,\ \Omega=\Omega_v}, \\
\beta(K) &=& \left.\varepsilon^\prime_\Omega(\Omega,K)\right|_{\Omega=\Omega_v},  \label{eq:beta(K)}
\end{eqnarray} 
and the primes denoting partial derivatives with respect to the corresponding variables, e.g., $\varepsilon^\prime_\Omega=\partial\varepsilon/\partial\Omega$. Here we have taken into account that $\varepsilon(\Omega_v,K_\parallel)=0$ (by definition of $\Omega_v$) and $\left.\varepsilon^\prime_{K_x}(\Omega,K)\right|_{K_x=0} = \left.\left[\left(K_x/K\right)\varepsilon^\prime_K\right]\right|_{K_x=0} = 0$.

The function $\zeta(\Omega,K_\parallel)$ in the upper vicinity of $\Omega=\Omega_v(K_\parallel)$ is then
\begin{eqnarray}
\zeta(\Omega,K_\parallel) &\approx& \frac{1}{2}+\frac{1}{2\pi K_\parallel}\int_{-\infty}^{+\infty}{\frac{dK_x}{\left(1+K_x^2/K_\parallel^2\right)\left[\beta(K_\parallel)(\Omega-\Omega_v) + \alpha(K_\parallel) K_x^2 \right]}} \nonumber \\
&=& \frac{1}{2} + \frac{1}{2K_\parallel\sqrt{\alpha(K_\parallel)\beta(K_\parallel)}}\frac{1}{\sqrt{\Omega-\Omega_v}}. \label{eq:zeta_approx}
\end{eqnarray}
Here we neglected $O[(\Omega-\Omega_v)^2]$ (as we are considering a small vicinity of $\Omega=\Omega_v$), $O(K_x^4)$ (as the main contribution into the integral is from $K_x\approx0$), and $O[(\Omega-\Omega_v)K_x^2]$ (due to the combination of the above two reasons). 
Similarly, in the upper vicinity of $\Omega=\Omega_v(K_\parallel)$ we obtain
\begin{eqnarray}
\int_{-\infty}^{+\infty}{dK_x^\prime\frac{I(\Omega,\mathbf{K^\prime})}{{K^\prime}^2\varepsilon(\Omega,K^\prime)}} \approx \frac{\pi \left.I(\Omega_v,\mathbf{K})\right|_{K_x=0}}{\sqrt{\alpha(K_\parallel)\beta(K_\parallel)} K_\parallel^2 \sqrt{\Omega-\Omega_v}},   \label{eq:I_approx}
\end{eqnarray}
where we have assumed that the function $I(\Omega,\mathbf{K})$ varies slowly near the point $\Omega=\Omega_v$, and does not have branch cuts.

Combining (\ref{eq:zeta_approx}) and (\ref{eq:I_approx}) in (\ref{eq:rho(w,k)}), we obtain the approximation (\ref{eq:rho_approx}) for $\tilde\rho(\Omega,\mathbf{K})$ in the upper vicinity of $\Omega=\Omega_v$.

\bibliographystyle{jpp}

\end{document}